\documentclass[twocolumn,pre,showpacs]{revtex4}

\usepackage{graphicx}
\usepackage{dcolumn}
\usepackage{bm}

\begin{document}

\preprint{APS/123-QED}

\title{Transient Turbulence in Taylor-Couette Flow}

\author{Daniel Borrero-Echeverry}
\affiliation{
Center for Nonlinear Science and School of Physics, Georgia Institute of Technology, Atlanta, Georgia 30332-0430, USA}
\author{Randall Tagg}
\affiliation{Department of Physics, University of Colorado, Denver, CO 80217-3364, USA}
\author{Michael F. Schatz}
\affiliation{
Center for Nonlinear Science and School of Physics, Georgia Institute of Technology, Atlanta, Georgia 30332-0430, USA}

\date{\today}

\begin{abstract}
Recent studies have brought into question the view that at sufficiently high Reynolds number turbulence is an asymptotic state. We present the first direct observation of the decay of turbulent states in Taylor-Couette flow with lifetimes spanning five orders of magnitude. We also show that there is a regime where Taylor-Couette flow shares many of the decay characteristics observed in other shear flows, including Poisson statistics and the coexistence of laminar and turbulent patches. Our data suggest that characteristic decay times increase super-exponentially with increasing Reynolds number but remain bounded in agreement with the most recent data from pipe flow and with a recent theoretical model. This suggests that, contrary to the prevailing view, turbulence in linearly stable shear flows may be generically transient.
\end{abstract}

\maketitle

Given its wide applicability and importance, it is somewhat surprising that there is still a rather incomplete understanding of the physics underlying turbulence. While the statistical approach of the past century has provided great insight, it has proven to have limited quantitative predictive power, even for the simplest of flows. Dynamical systems has emerged as a complementary technique and has shown promise in clarifying the physics of turbulence \cite{Eckhardt2008a}. In this picture, turbulent flows are viewed as trajectories in a high-dimensional phase space that wander between unstable solutions that coexist with the laminar solution \cite{Hof2004}.

Experiments in shear flows have shown that below the onset of turbulence, but at sufficiently high Reynolds number ($Re = UL/\nu$, where $U\mathrm{,}\,L \mathrm{,}\,\mathrm{and}\,\nu$ are the velocity and length scales of the problem and the kinematic viscosity of the fluid, respectively), states where well-defined regions of turbulent and laminar flow coexist can be reached by finite-amplitude perturbations to the laminar flow. For a range of $Re$ these states decay back to the laminar state. The lifetime of an individual event is very sensitive to the details of the initial perturbation \cite{Faisst2004,Schneider2008}. However, ensembles of experiments using similar perturbations have been found to have an exponential distribution of lifetimes for fixed $Re$ \cite{Bottin1998,Peixinho2006}. This distribution is consistent with a chaotic repeller model \cite{Kadanoff1984,Kantz1985}. Some numerical simulations and experiments have concluded that at moderate $Re$s the characteristic lifetimes of these states increase with increasing $Re$ until at some critical Reynolds number $Re_c$ they diverge and turbulence becomes sustained \cite{Daviaud1992,Bottin1998,Peixinho2006,Willis2007}. Above $Re_c$ the dynamics of the turbulent state are typically associated with those of a chaotic attractor \cite{Eckhardt2008}. Other experiments and simulations suggest that characteristic lifetimes do not diverge, instead increasing very rapidly but remaining finite at finite $Re$ \cite{Hof2006,Lagha2007,Hof2008,Manneville2008}. This implies that turbulence is not a permanent state of the flow for \emph{any} $Re$ and is instead generically transient, if very long-lived. This result has important repercussions when one considers the possibility of controlling the transition to/from turbulence by exploiting dynamical connections between the laminar and turbulent states \cite{Shinbrot1993}. The issue remains a subject of debate in part because the open flows (i.e., pipe flow and plane Couette flow) have been used to address this question. This limits the maximum time during which a turbulent episode may be observed and introduces complications associated with inlet conditions. Numerical exploration of this question is also problematic since it requires many highly-resolved simulations over long times to obtain significant statistics. 

\begin{figure}[b]
\includegraphics[width=\columnwidth]{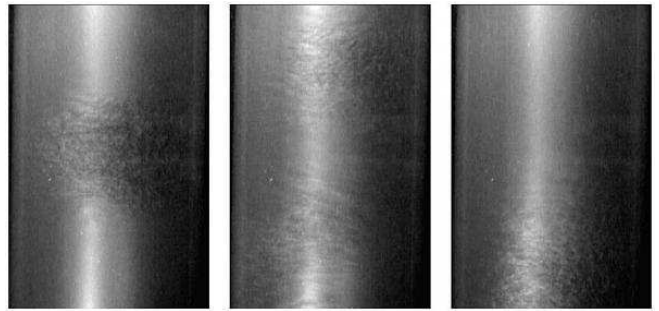}
\caption{\label{fig:turbpics}Photographs of turbulent patches in TCF at $Re$ = 7500 with only the outer cylinder rotating. In this regime, turbulent patches coexist with the laminar flow and evolve in space and time. For all $Re$ studied, these patches decay away in a probabilistic manner with a characteristic timescale $\tau$ dependent on $Re$. The photographs show a 25 cm high region of the flow.}
\end{figure}

We present the first measurements of lifetimes of transitional states in Taylor-Couette flow (TCF). TCF (i.e., the flow between rotating, concentric cylinders) is most famous for a transition to turbulence preceded by a hierarchy of bifurcations that arise from centrifugal instabilities \cite{Andereck1986,Coles1965}. However, TCF can bypass these instabilities and make a direct transition to turbulence. Several studies have looked at transitional states that arise in certain regimes of counter-rotating TCF \cite{Colovas1997,Litschke1998}, but little work has been done when only the outer cylinder is allowed to rotate. We believe that this regime of TCF is ideally suited to the experimental investigation of the lifetimes of transitional states. 

In this regime, the transition to turbulence is abrupt and is characterized by spatially and temporally intermittent patches of turbulence that coexist with a laminar background \cite{Coles1965} (see Fig. \ref{fig:turbpics}). Transition is observed to occur despite calculations that show that circular Couette flow should be linearly stable for all outer cylinder rotation rates as long as the inner cylinder is stationary \cite{Esser1996}. These are features that are observed in other transitional shear flows \cite{Tillmark1992,Wygnanski1973}. However, because TCF is streamwise periodic, it allows for arbitrarily long observation times, while avoiding the problems of contamination from the inlet that plague plane Couette and pipe facilities. Furthermore, using this untested geometry allows us to test the generality of the trends observed by earlier studies of the decay times of transitional states. 

Our experimental apparatus was a vertical Taylor-Couette system with a glass outer cylinder of radius $r_o$ = 7.620$\pm$0.008 cm and a brass inner cylinder of radius $r_i$ = 6.638$\pm$0.001 cm. The gap width $d = r_o-r_i$ was 0.982 cm and the radius ratio $\eta = r_i/r_o$ was 0.871. The cylinder ends were attached to the outer cylinder and the cross-stream aspect ratio $\Gamma = L/d$ was 36. Each cylinder was driven by a stepper motor and their angular velocities $\omega_{o,i}$ could be set to better than $0.5\%$. The entire apparatus was submerged in a temperature bath set to $20.00^{\circ}$C that varied by less than $\pm0.05^{\circ}$C over several days.

Our working fluid was distilled water with 2.2\% Kalliroscope AQ 1000 added by volume for flow visualization \cite{Matisse1984}. Adding Kalliroscope to water slightly changes its viscosity, so we measured it using a Cannon-Fenske routine viscometer. We found that it had a kinematic viscosity $\nu$ of $1.0298\pm 0.0025\,\mathrm{mm}^2{\mathrm{/s}}$  at $20^{\circ}$C. We also studied the rheology of the suspension using a stress-controlled rheometer and found no deviations from Newtonian behavior. These results, along with the geometric characteristics of the apparatus, allowed us to set the outer cylinder $Re$ to within 1\%. Here,
\begin{eqnarray}
Re = \frac{U L}{\nu} = \frac{r_o\, \omega_o\,(r_o-r_i)}{\nu}.
\label{eq:Reynoldsnumber}
\end{eqnarray}

\begin{figure}
\includegraphics[width=\columnwidth]{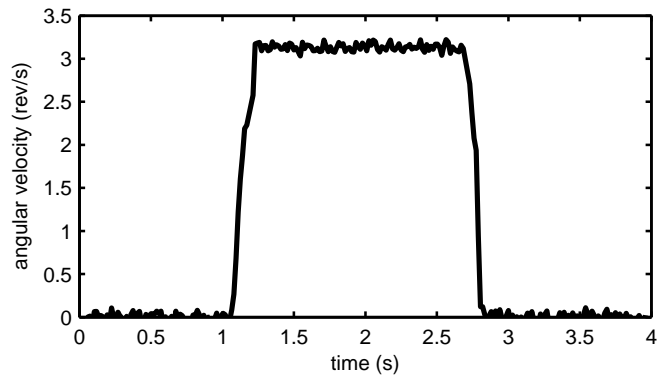}
\caption{\label{fig:perturbation} The inner cylinder angular velocity as a function of time shown above was measured using solid state rotation sensors. Experiments conducted at $Re = 7647$ showed measured lifetimes to be insensitive to variations in inner cylinder acceleration, maximum velocity, and pulse duration so the profile above was used for experiments at all values of $Re$ studied.}
\end{figure}
Like plane Couette and pipe flows, TCF (in the regime described above) requires a finite-amplitude perturbation to trigger transition. The laminar state was first prepared by accelerating the outer cylinder to the desired $Re$. Since at sufficiently high $Re$ ($\sim$16,000, in our case) ambient noise is sufficient to cause spontaneous transition, we limited our study to $Re$s significantly below this threshold. The system was allowed to run in Couette flow for several minutes ($\sim$ 3 radial diffusion times $t_d = d^2/\nu$) in order to eliminate any transients.

The flow was then perturbed by rapidly accelerating the inner cylinder in the direction opposite the rotation of the outer cylinder and immediately stopping it. Figure \ref{fig:perturbation} shows the angular velocity of the inner cylinder as a function of time. It is important to note that our perturbation is qualitatively different from those used in the pipe flow experiments \cite{Peixinho2006,Hof2006} and is similar to the quenching experiments conducted in plane Couette flow \cite{Bottin1998} because it is not a small, localized perturbation but a large, global one where the whole flow is disturbed. After briefly exhibiting featureless turbulence, the flow relaxed to an intermittent state like the one shown in Fig. \ref{fig:turbpics}. The flow was then observed until it relaminarized.

A CCD camera captured video of the flow at a resolution of 320 x 240 pixels. Except for the three highest Reynolds numbers studied, the video was captured at 30 frames per second. At these Reynolds numbers the frame rate was reduced to 5 frames per second due to data storage limitations since individual events could last many hours. The low resolution of the video stream allowed it to be analyzed in real-time by a computer. 

First, the video stream was separated into ten frame segments and a frame from each segment was subtracted from a reference frame in the segment preceding it. The resulting image was thresholded to highlight only pixels that were significantly different from the same pixel in the reference image to reduce noise. The elements of this binary image where added together and the resulting number $N$ was compared to an empirically determined threshold, $N_T$. If $N$ dropped below $N_T$ for 30 seconds (turbulent patches were never observed to return after disappearing for more than a few of seconds), the system stopped acquiring data and prepared for the next run. The video data was saved and the accuracy of the automated lifetime measurements was verified by hand for a random selection of the experimental trials.

\begin{figure}
\includegraphics[width=\columnwidth]{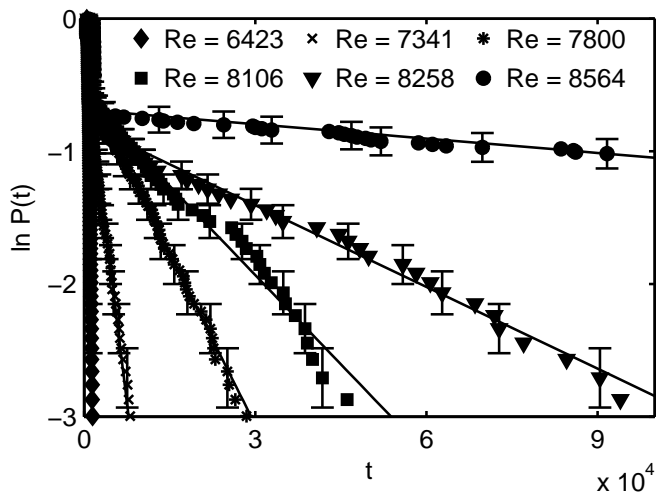}
\caption{\label{fig:probability} The probability of observing events whose lifetime is longer than time t decreases exponentially for long times with a decay constant that increases with increasing $Re$. Times are expressed in dimensionless time units $t_{nd} = d/r_o\, \omega_o$. Only a fraction of the total events used to obtain the results in Fig. \ref{fig:lifetimes} are shown for clarity. The error bars indicate sampling error and are shown only at representative points. The solid lines show the weighted least squares linear fits to the tails of the distributions.}
\end{figure}
After repeating the experiment many times (between 200 and 1200 times depending on $Re$) at fixed $Re$, we calculated the cumulative probability distributions $P(t,Re)$ shown in Fig. \ref{fig:probability}. $P(t,Re)$ is the probability of observing turbulence at some time $t$ after the initial perturbation for a fixed value of $Re$.

These curves have two salient features. First, the distributions have exponentially decaying tails (i.e., $P(t,Re) \sim exp(-(t-t_0)/\tau(Re))$, where $t_0$ is the time associated with the initial formation of the turbulent state and $\tau(Re)$ is a decay constant that depends on $Re$). This indicates that for long times the decay of turbulence is a Poisson process, which is a hallmark of a chaotic repeller \cite{Kadanoff1984,Kantz1985}. The same behavior has been reported in all previous studies of the decay of turbulence in plane Couette and pipe flows \cite{Peixinho2006,Hof2006,Bottin1998,Hof2008,Lagha2007,Manneville2008,Willis2007}. $\tau$ was calculated by fitting a straight line to the tail of each distribution. As a preliminary test of the role played by the end walls in the decay, we checked 150 long-lived events at various $Re$s to see if turbulent patches were last seen at the top, middle, or bottom of the cylinder. We found that each location was equally likely. A more thorough study is underway to check the sensitivity of decay times to varying end conditions (e.g., variations in $\Gamma$). 

Second, the curves show that a significant fraction ($\sim$ 30\%) of the experiments relaminarized immediately after the perturbation, almost independently of $Re$. Because our perturbation consists of globally disorganizing the fluid, one can think of it as moving the system away from the laminar attractor in a random direction in phase space. If the perturbation puts the system in the basin of attraction of the turbulent state, the system will remain turbulent. If the perturbation leaves the system in the basin of attraction of the laminar state, the turbulence decays immediately. While the short-lived events where not used in determining the decay constants, the duration of these events allowed us to estimate $t_0$. This was determined to be $\sim$ 120 nondimensional time units.

\begin{figure}
\includegraphics[width=\columnwidth]{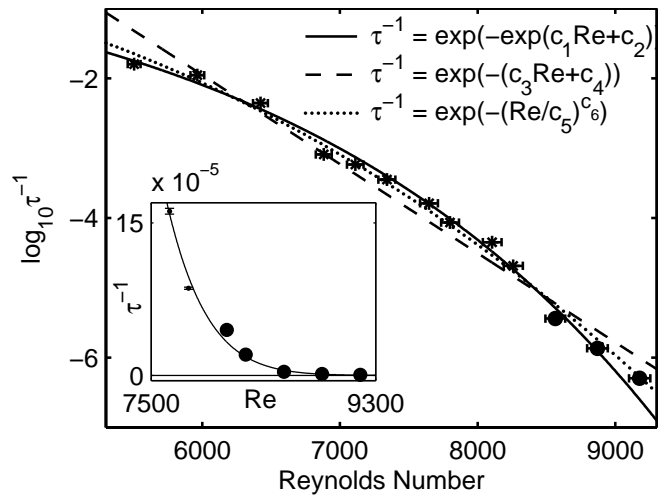}
\caption{\label{fig:lifetimes} Characteristic lifetimes as a function of $Re$ in dimensionless time units. $\ast$ indicates data sets for which all the decays were directly observed. The points indicated with $\bullet$ correspond to data sets for which a maximum observation time was set as indicated in the text. Horizontal error bars represent absolute error in $Re$ based on experimental limitations. Errors in log$_{10}\,\tau^{-1}$ are on the order of the symbol size and were estimated by the uncertainty calculated for the slope of the linear fits shown in Fig. \ref{fig:probability}. Inset: A plot of $\tau^{-1}$ vs. $Re$ on a linear scale shows that $\tau^{-1} \rightarrow 0$ only as $Re \rightarrow \infty$.}
\end{figure}

We observed that the characteristic lifetimes increased very rapidly with $Re$, but did not find a $Re$ for which we could not observe decay. At $Re$ = 9,922, we observed an event that lasted 29 hours before relaminarizing. The lifetimes became so long that we imposed maximum observation times for the highest Reynolds numbers studied. Because the experiments at lower Reynolds numbers were conducted first, the maximum observation times were chosen by linearly extrapolating from the decay constants calculated for the previous three experiments on a plot of log$_{10}\,\tau(Re)$ vs. $Re$ and multiplying this time by 1.5. This prevented us from directly observing the decays of the longest-lived events, but still left enough events of intermediate length to get good estimates of the characteristic lifetimes. To validate this procedure, we used it to measure the characteristic lifetime for $Re$ = 7,111. The resulting lifetime differed from that measured by observing all events by less than 1\%. 

We also conducted five experimental runs of 300 events each at $Re = 7647$ to check the sensitivity of our results to the details of the perturbation. Increases in the pulse duration of up to a factor of 10 did not change our results significantly. Neither did decreasing the inner cylinder acceleration by a factor of 60. The measured lifetimes were also insensitive to the maximum inner cylinder velocity as long as this was on the order of the speed of the outer cylinder. However, if the maximum speed was much lower than that, the fraction of events that relaminarized immediately increase dramatically. This agrees with recent experiments in pipe flow \cite{Lozar2009} that showed that as long as the perturbation is large enough to cause the transition to turbulence the observed decay times do not depend on the details of the perturbation.

Figure \ref{fig:lifetimes} shows the characteristic lifetimes $\tau(Re)$ as a function of $Re$. We fit the data with the various functional forms suggested by Hof et al. \cite{Hof2008} and found that
\begin{eqnarray}
\tau(Re)^{-1} = exp(-exp(c_1\,Re+c_2)),
\label{eq:tauvsRe}
\end{eqnarray}
with $c_1 = 3.61 \times 10^{-4}$ and $c_2 = -0.59$ best captured the trend (i.e., had the smallest residuals). While Eq. \ref{eq:tauvsRe} seems to fit the observed trend, goodness-of-fit statistics indicate that it is not statistically significant fit (i.e., $\chi_\nu^2\,\gg\,1)$. Therefore, we do not claim that it represents the actual functional dependence of $\tau$ on $Re$, but only that lifetimes grow faster than exponentially but remain bounded. However, it should be noted that the scaling of Eq. \ref{eq:tauvsRe} agrees with the most recent results for pipe flow \cite{Hof2008} and with the only theoretical prediction of $\tau(Re)$ \cite{Goldenfeld2009}. As shown in Fig. \ref{fig:lifetimes}, alternative fits to the data are possible (e.g., $\tau^{-1} = exp(-(Re/c_5)^{c_6})$ with $c_5 = 3305$ and $c_6 = 2.62$) and differentiating between them requires many more decades of data.

We have shown that when only the outer cylinder is allowed to rotate, transitional states in Taylor-Couette flow share many of the decay characteristics observed in other canonical shear flows. We showed that the decay of these states is a Poisson process, which suggests that the correct model for turbulence in this regime is a chaotic repeller. Furthermore, we have shown through direct observation that characteristic decay times increase super-exponentially with increasing $Re$ but remain bounded, in agreement with the most recent data from pipe flow \cite{Hof2008} and with a recent theoretical prediciton for scaling \cite{Goldenfeld2009}. This suggests that contrary to the conventional view, turbulence may be generically transient for all $Re$ in linearly stable shear flows. 

However, several open questions remain. As has been pointed out recently, it may be possible that turbulence can become sustained globally even if it is transient locally in a manner similar to directed percolation \cite{Manneville2009}. This prediction comes from by a highly simplified model of plane Couette flow and must be verified experimentally for real fluid flows since it would be very computationally expensive to address in direct numerical simulations. Other researchers have pointed out that the laminar state of finite-sized Taylor-Couette system is nontrivial \cite{Andereck1986} and the role played by end-effects (e.g., Ekman pumping) in relaminarization is still unknown. Finally, it is unclear what effects curvature and rotation have on the decay of turbulence. Work is currently underway to address these issues. 

The authors would like to thank A. Fernandez-Nieves and B. Sierra-Mart\'in for performing rheoscopic measurements of our Kalliroscope suspensions and T. Mullin and A. De L\'ozar for helpful discussions.

\bibliography{turbulence}

\end{document}